\documentclass[aip,reprint,amsmath, twocolumn, groupedaddress]{revtex4-1}
\usepackage[T1]{fontenc}
\usepackage[latin9]{inputenc}
\usepackage{amstext}
\usepackage{graphicx}
\usepackage{amssymb}
\usepackage{color}

\definecolor{red}{RGB}{255, 0,0}
\definecolor{green}{RGB}{0, 128, 0}
\definecolor{blue}{RGB}{0, 0,  113}

\makeatletter

\providecommand{\LyX}{L\kern-.1667em\lower.25em\hbox{Y}\kern-.125emX\@}

\makeatother

\begin{document}

\preprint{}

\title{Fluctuations From Edge Defects in Superconducting Resonators}

\author{ C. Neill$^1$}
\author{ A. Megrant$^1$}
\author{ R. Barends$^1$}
\author{ Yu Chen$^1$}
\author{ B. Chiaro$^1$}
\author{ J. Kelly$^1$}
\author{J. Y. Mutus$^1$}
\author{ P. J. J. O'Malley$^1$}
\author{ D. Sank$^1$}
\author{ J. Wenner$^1$}
\author{ T. C. White$^1$}
\author{ Yi Yin$^1$}
\altaffiliation[Present address: ]{Department of Physics, Zhejiang University, Hangzhou 310027, China}
\author{ A. N. Cleland$^{1,2}$}
\author{ John M. Martinis$^{1,2}$}
\email{martinis@physics.ucsb.edu}

\affiliation{$^1$Department of Physics, University\,of\,California, Santa\,Barbara, CA 93106, USA}
\affiliation{$^2$California NanoSystems Institute, University\,of\,California, Santa\,Barbara, CA\,93106, USA}

\date{\today}

\begin{abstract}
Superconducting resonators, used in astronomy and quantum computation, couple strongly to microscopic two-level defects.  We monitor the microwave response of superconducting resonators and observe fluctuations in dissipation and resonance frequency.  We present a unified model where the observed dissipative and dispersive effects can be explained as originating from a bath of fluctuating two-level systems.  From these measurements, we quantify the number and distribution of the defects. 
\end{abstract}

\maketitle


Superconducting microwave resonators are solid state devices that play important roles in quantum computation as memory elements, zeroing registers and data transfer buses \cite{galiautdinov2012resonator, mariantoni2011implementing}. Resonators are also used as pixels in submillimeter-wavelength, energy-resolved telescopes where they are known as microwave kinetic inductance detectors (MKIDs) \cite{zmuidzinas2003broadband, mazin2012superconducting}. The performance of these devices is limited by bistable tunneling defects, which are present in the amorphous dielectrics on the device surface.  Originally studied in the context of glasses \cite{phillips1999two}, these microscopic two-level systems (TLSs) reduce MKID sensitivity through the introduction of noise \cite{gao2008experimental, barends2010reduced, gao2008semiempirical, gao2011strongly} and limit quantum coherence through the absorption of energy \cite{martinis2005decoherence, kline2008josephson}.

Previous studies have shown that TLSs cause fluctuations in the resonance frequency of superconducting resonators \cite{vissers2012reduced, barends2010reduced, gao2008semiempirical}.  The amplitude of these fluctuations decreases with increasing resonator excitation energy  \cite{gao2008semiempirical}, such that when exciting resonators with energies exceeding $10^6$ photons, the excess dissipation noise can be reduced to below the quantum limit \cite{gao2011strongly}.  However, when operated at small excitation energies, such as in quantum information applications, these TLS fluctuators present a fundamental challenge \cite{corcoles2011protecting}.

In this Letter, we present measurements of fluctuations in both the resonance frequency and the internal loss of superconducting microwave resonators, driven with excitation energies ranging from a single photon to $10^4$ photons.  We describe a model of fluctuating TLSs that allows us to extract the number and distribution of these defects.  Surprisingly, we find that, despite the macroscopic size of the devices, the observed fluctuations are dominated by a small number of defects located near the metal edges.

\begin{figure}[b]
\includegraphics[scale=0.43]{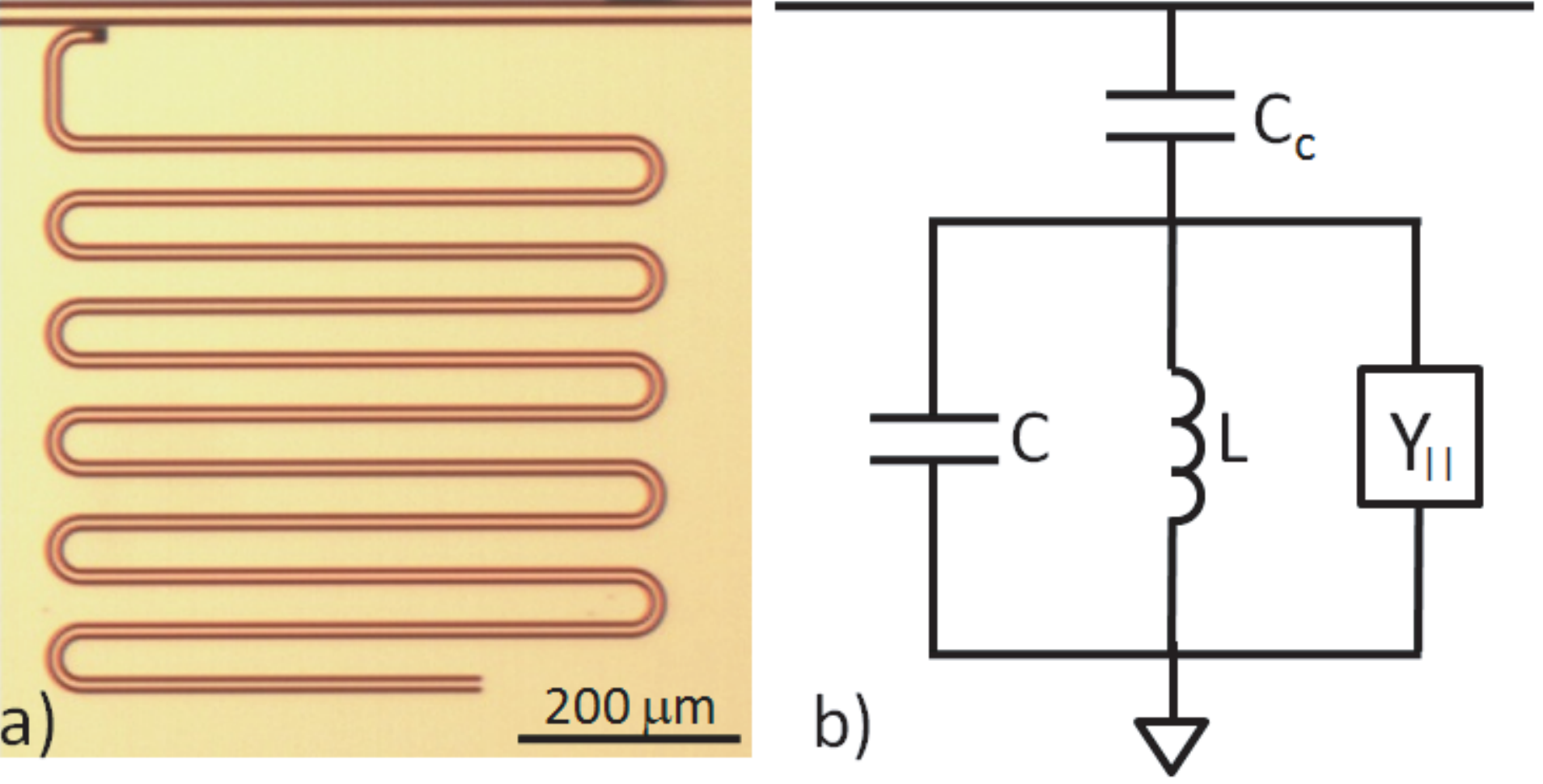}
\caption{
(a)  Optical micrograph of a quarter wavelength CPW resonator used in this experiment.  Light regions correspond to aluminum; dark regions correspond to exposed sapphire.
(b) Equivalent circuit for the CPW near its fundamental resonance frequency, with lumped capacitance $C$, inductance $L$, and coupling capacitance $C_c$ to the measurement transmission line. Loss and fluctuations in the dielectric response of the resonator are captured by the equivalent parallel admittance $Y_{||}$, which is found to vary with microwave drive power and time. }
\label{fig:device}
\end{figure}

The devices used in this experiment were quarter-wavelength coplanar waveguide (CPW) resonators capacitively coupled to a transmission line; such a device is shown in Fig.\,\ref{fig:device}.  The resonators were patterned from a 100 nm thick aluminum film atop a 600 $\mu$m thick sapphire substrate.  Each $6 \times 6$ mm$^2$ die contained 12 resonators connected to a single transmission line. The CPW center strips were 6 $\mu$m wide with 4 $\mu$m spacing to ground, and were approximately 6 mm in length with distinct resonance frequencies ranging from 4.88 to 5.03 GHz. An individual die was wire bonded into an aluminum sample mount and placed on the 50 mK stage of an adiabatic demagnetization refrigerator.  Details regarding attenuators, filters, shielding, and amplifiers in the experimental setup, as well as sample growth conditions, are available in Refs. \onlinecite{megrant2012planar, barends2011minimizing}.

\begin{figure}
\includegraphics[scale=0.55]{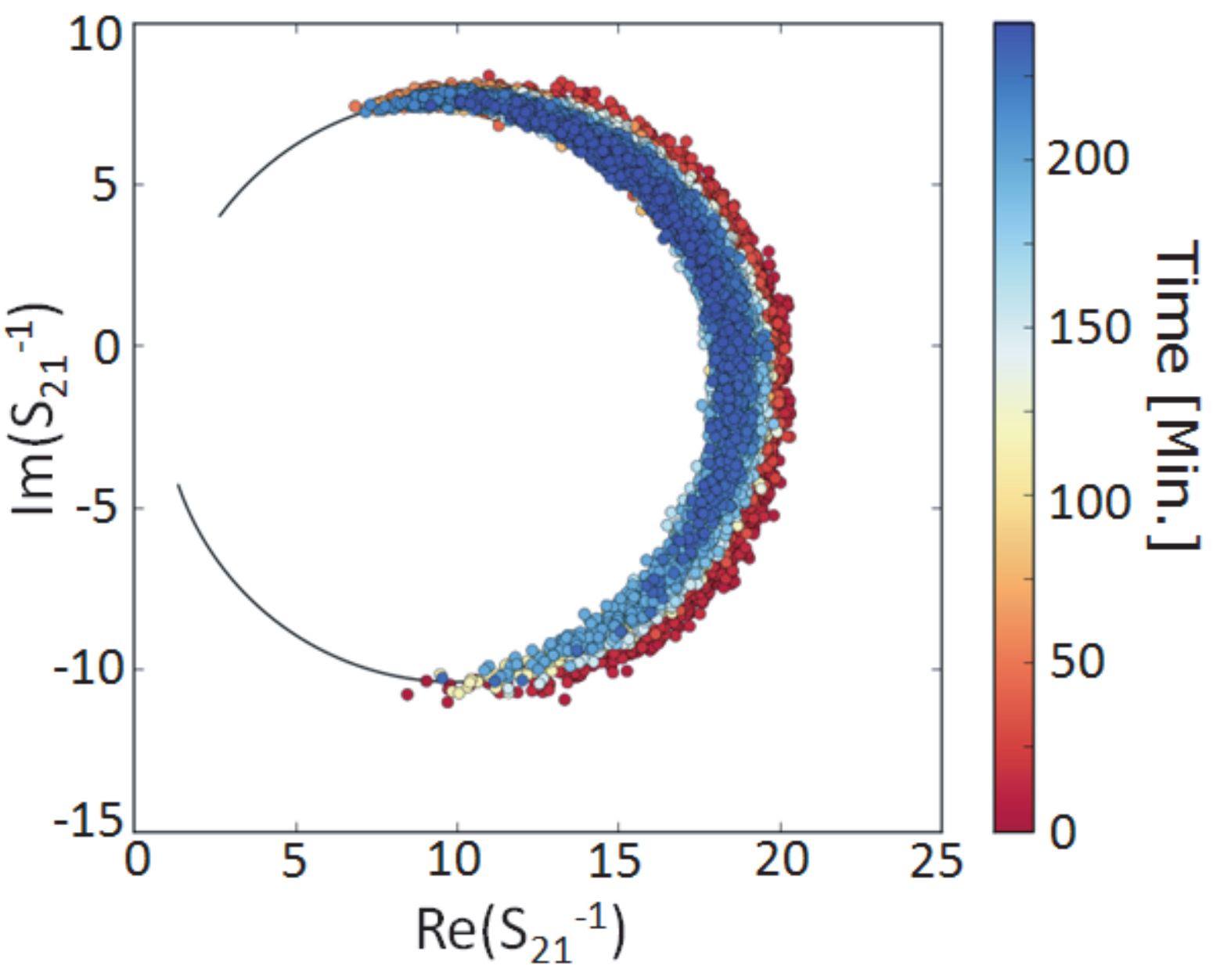}
\caption{
Fixed-frequency measurements of the complex transmission amplitude $S_{21}^{-1}$ versus time (color). The resonance circle, shown in black, is traced over as the resonator response fluctuates in time.
}
\label{fig:transmission}
\end{figure}

Near the CPW fundamental resonance frequency, the resonator can be represented by the equivalent circuit shown in Fig.\,\ref{fig:device}(b), using a lumped capacitance $C$ and parallel inductance $L$, with resonator loss and fluctuations in the resonator's response accounted for by the parallel admittance $Y_{||}$. The average resonator frequency is $\overline{\nu}_R = 1/2 \pi \sqrt{LC}$; the resonator characteristic admittance is $Y_0 = \sqrt{C/L}$. For drive frequencies $\nu$ near $\overline{\nu}_R$, the resonator admittance $Y(\nu)$ can be written in dimensionless form as
\begin{equation}\label{eq.ydimen}
    y(\nu) \equiv \frac{Y(\nu)}{Y_0}  \approx \frac{1}{Q_i} + 2 i \frac{\nu-\nu_R}{\overline{\nu}_R},
\end{equation}
with real part $y_R$ equal to the internal resonator loss $1/Q_i$ and imaginary part $y_I$ given by the fractional detuning $2 (\nu-\nu_R) / \overline{\nu}_R$ of the microwave drive frequency $\nu$ from the instantaneous value of the resonance frequency $\nu_R$.

The devices were characterized using a vector network analyzer to measure the transmission scattering amplitude $S_{21}$ through the coupled transmission line. The inverse scattering amplitude of the CPW is given in terms of $y(\nu)$ by \cite{megrant2012planar}
\begin{equation}\label{eq.scatter}
    S_{21}^{-1}(\nu) = 1 + \frac {1}{Q_c  e^{-i \phi}} \frac{1}{y},
\end{equation}
where $Q_c$ is the resonator-transmission line coupling quality factor and $\phi$ is a geometry-dependent phase shift resulting from impedance mismatches on either side of the transmission line.

\begin{figure}
\includegraphics[scale=0.55]{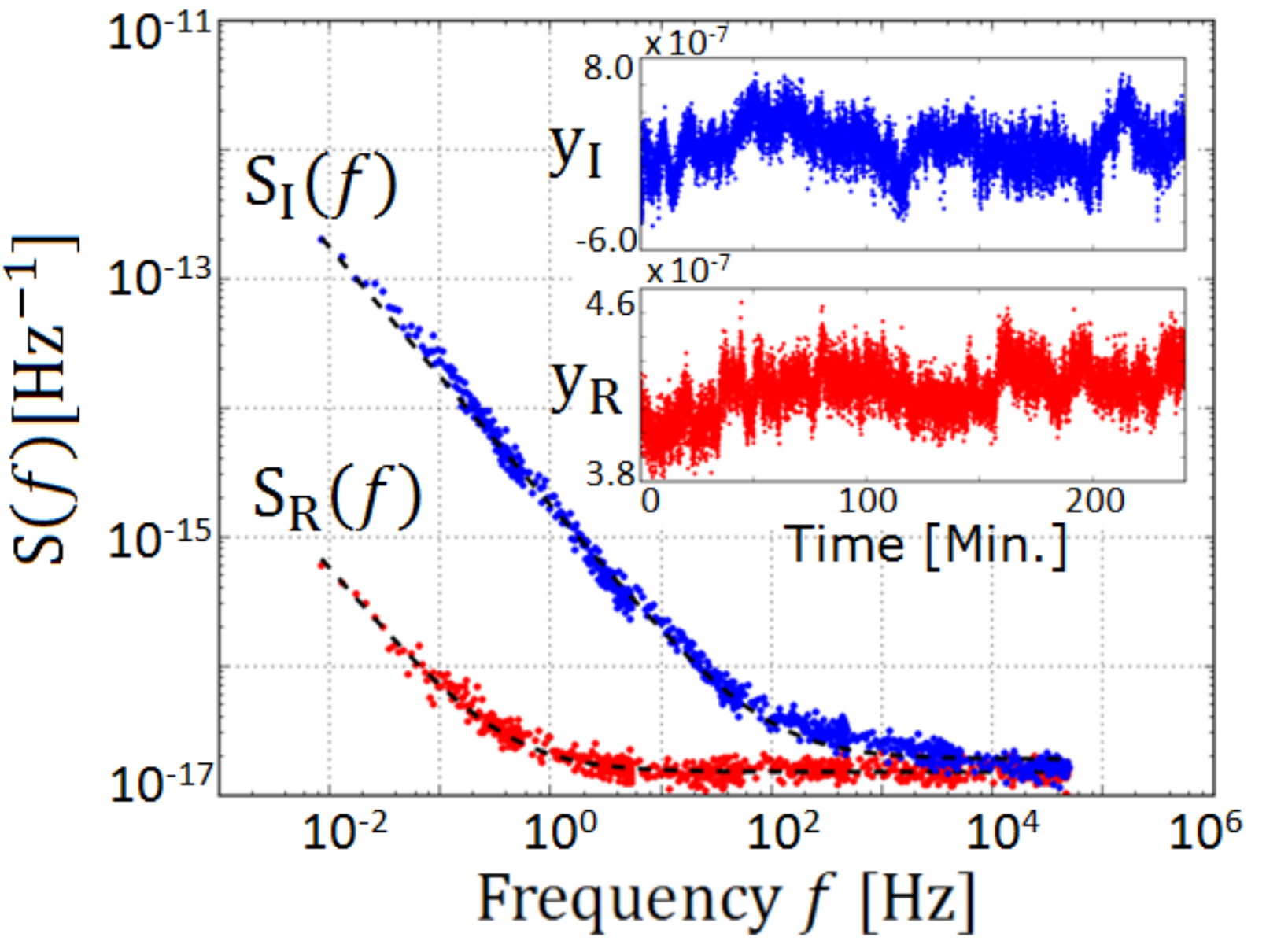}
\caption{
 Inset: Time-dependent real and imaginary parts of the dimensionless admittance $y$, corresponding to internal resonator loss and fractional frequency detuning, respectively. Main panel: Power spectral densities $S_R(f)$ and $S_I(f)$ of the real and imaginary parts of $y$, as a function of Fourier transform frequency $f$.  Dashed lines are fits to $\alpha/f + \beta$.
}
\label{fig:spectra}
\end{figure}

\begin{figure*}
\includegraphics[scale=0.45]{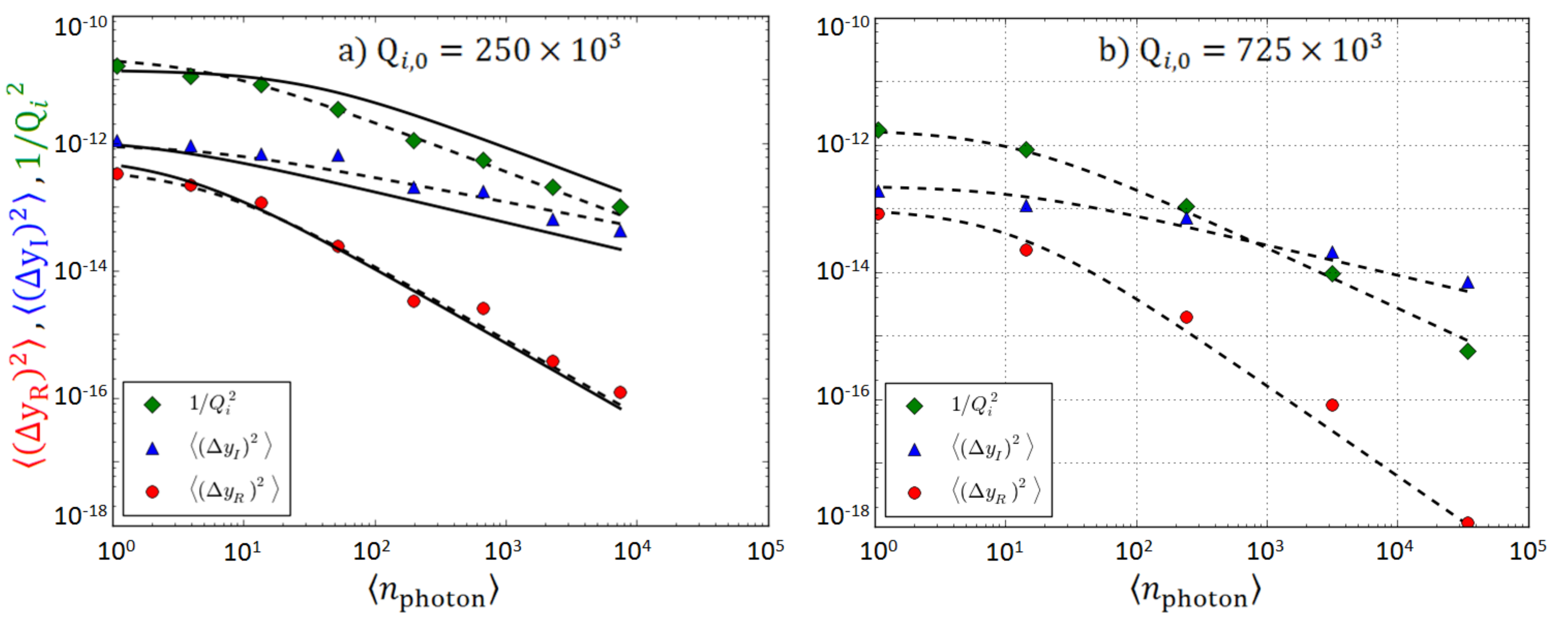}
\caption{
Power-dependent loss squared $1/Q_i^2$ (green) and variances in $y_R$ (red) and $y_I$ (blue), as a function of resonator power, plotted in units of resonator photon occupation number \cite{barends2009noise} for two different resonators.  (a) Resonator with low-power internal quality factor $Q_{i,0} = 250,000$, and (b) resonator with $Q_{i, 0} = 725,000$.  We observe lower noise for the device with lower loss.  The dashed lines overlaying the blue, green and red datasets are proportional to the measured dissipation, dissipation squared, and dissipation cubed, respectively (see text).   Solid lines correspond to electromagnetic simulations of the CPW loss and noise.}
\label{fig:processedData}
\end{figure*}

Here we focus on measurements taken at a \emph{fixed} microwave drive frequency $\nu$, near the resonance frequency $\overline{\nu}_R$, during which we capture the time-dependent value of $S_{21}^{-1}$.  Each data set comprises 16,000 measurements taken at a sampling rate ranging from 10 kHz to 1 Hz.  Single-frequency transmission data is shown in Fig.\,\ref{fig:transmission}, taken at 4.98 GHz with microwave drive power $P$ corresponding to an average resonator occupation of 7,500 photons. The response sweeps out a circle in the complex $S_{21}^{-1}$ plane, although we emphasize that these measurements are taken at a fixed frequency; the response varies due to fluctuations in the resonator parameters, which correspond to fluctuations in the parallel admittance $Y_{||}$ in the equivalent circuit.

The time-dependent components of $y$ can be determined from the measured $S_{21}^{-1}$ using Eq.\,(\ref{eq.scatter}), and are displayed in the inset of Fig.\,\ref{fig:spectra}.  We Fourier-transform the data to yield the power spectral densities $S_R(f)$ and $S_I(f)$ of $y_R$ and $y_I$ as a function of the Fourier-transform frequency $f$, plotted in the main panel of Fig.\,\ref{fig:spectra}. These spectral densities are calculated by dividing each time-domain data set into segments of 256 data points, followed by a Fourier transform and averaging in the frequency domain. We see that both power spectra $S_R(f)$ and $S_I(f)$ have a clear $1/f$ dependence, flattening out above a few Hz into a white noise background where the measurement noise dominates.  The cross spectral density of   $y_R$ and $y_I$ (not shown) indicates that the fluctuations in the two components are uncorrelated with one another. We note that $1/f$ noise originating in the microwave amplifier chain is far below the levels measured here.

Measurements of the type shown in Fig.\,\ref{fig:spectra} were made for different microwave drive powers $P$; in general, the level of fluctuations fall monotonically with drive power. Both the $1/f$ frequency dependence and the power dependence are consistent with lossy two-level systems interacting with the electric fields of the resonator, causing loss and dispersive frequency shifts. We therefore desire a detailed model to provide a more quantitative understanding of these effects.

The response of TLSs is well described by the Bloch equations \cite{phillips1999two, gao2008semiempirical}.  These equations describe an ensemble of TLSs with dipole moments $d$ and TLS decoherence rate $\Gamma_2 = 1/T_2$. The resulting linear response can be expressed as an equivalent circuit admittance $Y_{TLS}$, as in the equivalent circuit in Fig. \ref{fig:device}(b), with
\begin{equation}\label{eq:YTLSTheory}
    Y_{TLS} = G \frac{n d^2}{3 \epsilon \hbar}
        \sum_j \left ( \frac{\Gamma_2}{\Delta \omega_j^2 + \Gamma_2^2 \kappa^2} + i \, \frac{\Delta \omega_j}{\Delta \omega_j^2 + \Gamma_2^2 \kappa^2}  \right )
\end{equation}
where $G$ is an overall geometric scaling factor, $n$ is the number of TLSs per unit volume, $\epsilon$ is the permittivity of the dielectric in which the TLS are embedded, $\Delta \omega_j/2\pi$ is the detuning between the microwave drive frequency $\nu$ and the natural transition frequency of the $j$th TLS, and $\kappa = \sqrt{1 + E^2 / E_s^2}$ describes the saturation of the TLS in an electric field $E$, with saturation field $E_s$ \cite{Note1}. The local electric field varies with the geometric location of the TLS in the resonator, but for now we neglect this detail; we will return to this question below when we discuss a finite element model. Equations (\ref{eq.ydimen}) and (\ref{eq:YTLSTheory}) allow us to relate the microscopic TLS model to the experimentally-accessible resonator loss $1/Q_{i}$ and instantaneous frequency detuning $\nu - \nu_R$.

The average admittance at low temperatures is approximated by replacing the sum in Eq.\,(\ref{eq:YTLSTheory}) with an integral over detuning $\Delta \omega$ and a uniform TLS density of states $\rho(\Delta \omega)$, yielding \cite{martinis2005decoherence, gao2008semiempirical}
\begin{equation}\label{eq.avgrealy}
    \mbox{Re}\{Y_{TLS}\} = G \, \delta_i / \kappa,
\end{equation}
where $\delta_i = \pi \rho n d^2 / 3 \epsilon$ is the intrinsic loss tangent of the dielectric.

We calculate fluctuations in the real and imaginary parts of $Y_{TLS}$ by assuming that the coupling to each TLS fluctuates in time with a fractional variance $A^2$ which ranges between 0 and 1; justification for this assumption comes from a good fit to the data.  Integrating over the TLS population yields the magnitude of the mean-squared fluctuations, \begin{eqnarray}
    \left < (\Delta Y_{R, TLS})^2 \right> &=& \int {\hbar \rho \left (\frac{G A n d^2}{3 \epsilon \hbar}  \frac{\Gamma_2} {(\Delta \omega)^2 + \Gamma_2^2 \kappa^2} \right)^2 \mathrm{d} (\Delta \omega)  }\nonumber\\
    &=& G^2 \delta_i^2 \frac {A^2} {N} \frac {1} {\kappa^3}, \label{eq:VarReY}
\end{eqnarray}
and
\begin{eqnarray}
    \left< (\Delta Y_{I, TLS})^2 \right> &=& \int {\hbar \rho \left (\frac{G A n d^2}{3 \epsilon \hbar}  \frac{\Delta \omega} {(\Delta \omega)^2 + \Gamma_2^2 \kappa^2} \right)^2 \mathrm{d} (\Delta \omega)  }\nonumber\\
    &=& G^2 \delta_i^2 \frac {A^2} {N} \frac {1} {\kappa}, \label{eq:VarImY}
\end{eqnarray}
where $N = 2 \pi \Gamma_2 \hbar \rho$ is the effective number of TLS coupled to the device.

Equations (\ref{eq.avgrealy}), (\ref{eq:VarReY}) and (\ref{eq:VarImY}) make specific predictions for the relation between loss and fluctuations: The variances in the dissipative and dispersive fluctuations should scale with the square of the intrinsic loss $\delta_i$. Furthermore, the loss and variance in resonator frequency should scale with microwave power $P$ as $1/\kappa$, while the variance in loss should scale as $1/\kappa^3$.

We can compare these predictions with our measurements by calculating the variances in the real and imaginary parts of the measured dimensionless admittance,  $\left< (\Delta y_R)^2 \right>$ and  $\left< (\Delta y_I)^2 \right>$. These variances are calculated by integrating the $1/f$ component of the measured power spectral densities in Fig. \ref{fig:spectra}.  We approximate the full integrals by numerically cutting off the integrals below 1 mHz and above 5 kHz; the results only depend on the logarithmic ratio of these cutoffs. The calculated variances are plotted as a function of resonator photon occupation number in Fig.\,\ref{fig:processedData}.

We compare the measured dependence of the squared loss $1/Q_i^2$, and the variances $\left< (\Delta y_R)^2 \right>$ and  $\left< (\Delta y_I)^2 \right>$, with the model predictions.  The power-dependent loss $1/Q_i$ can be scaled to match well to the dispersive fluctuations in $y_I$. The power dependence of loss cubed, $1/Q_i^3$, is also seen to match well to the dissipative fluctuations in $y_R$, as expected (we note that the relative magnitude of the dispersive and dissipative variances differs from one another by a factor of 2 at low power).  We see that at single photon excitations, $\left< (\Delta y_R)^2 \right> \approx \left< (\Delta y_I)^2 \right> \approx (1/30)(1/Q_{i})^2$; the loss $y_R$ fluctuates by almost 20\% of its mean value. We observe similar behavior for a device with an internal quality factor that is roughly 3 times greater; this device displays roughly an order of magnitude lower variance, in agreement with the model, which scales as $\delta_i^2$.

Assuming that the model variance $A^2$ is of order 1, Eq.\,(\ref{eq:VarReY}) and the measured fractional variance imply that a relatively small equivalent number of TLSs, approximately 30, are affecting the device performance. Smaller values of $A$ would imply fewer TLSs: as we do not yet observe single fluctuators, $A^2 \approx 1$ seems to be a reasonable estimate.

The contribution of different TLSs to the overall admittance $Y$ is weighted by the local electric field, which we ignored in our previous TLS model. To better understand this dependence, we used a finite element electromagnetic simulation to calculate the TLS contribution as a function of TLS location and local $\kappa(E)$. We model a thin (3 nm thick) uniformly lossy dielectric with a relative permittivity of $\epsilon = 10$ on all of the device interfaces (see Ref.\,\onlinecite{wenner2011surface}). Using the simulated electric fields, we calculate the percentage contribution to dissipation and noise as a function of distance from the substrate-electrode corners, shown in Table 1. The contribution to the overall loss from a small volume scales as $E^2$, and the local field $E$ scales with distance $x$ from the edge of a metal film as $1/\sqrt{x}$, resulting in loss contributions that are distributed logarithmically. The contribution to the overall variance, however, scales with the square of the loss, i.e. $E^4$, resulting in contributions dominated by the corners.

\begin{table}[t]
\begin{center}
\begin{tabular} { | c | c | c | c | c | c | }
	\hline
	 & 0 - 10 nm & 10 - 100 nm & .1 - 1 $\mu$m & 1 - 10 $\mu$m \\ \hline
	Loss & 29\% & 28\% & 30\% & 13\% \\ \hline
	Noise & 88\% & 11\% & 1\% & <1\% \\ \hline
\end{tabular}
\end{center}
\caption{
We simulate a CPW with a thin dielectric layer on all surfaces of thickness 3 nm and relative permittivity of 10. The percent contribution to loss and noise are presented as a function of distance from the metal-substrate corners.
}
\end{table}

In Fig.\,\ref{fig:processedData}(a) we plot the simulated power dependence of the loss squared, $1/Q_i^2$, and fluctuations in $Y_{R,TLS}$, $Y_{I,TLS}$, as solid lines overlaying the experimental data.  Closest agreement between data and simulation is seen by setting the TLS saturation field to 10.0 V/m, the intrinsic loss tangent to $\delta_i = 1.1 \times 10^{-3}$ and the TLS density to $2 /\mu$m$^3$.  These values agree with previous measurements \cite{shalibo2010lifetime, martinis2005decoherence, wang2009improving} and the resulting power dependence and number of TLSs are in good agreement with the data.

In conclusion, we have shown measurements of power-dependent fluctuations in the loss and resonance frequency of superconducting resonators.   We have presented a model that reproduces the measured power dependence and scaling with internal loss.   From these measurements we were able to estimate the number of defects that contributed to the resonator fluctuations.  The response of these devices appear to be dominated by a few dozen TLSs.

\begin{acknowledgments}
Devices were made at the UC Santa Barbara Nanofabrication Facility, a part of the NSF-funded National Nanotechnology Infrastructure Network. This research was funded by the Office of the Director of National Intelligence (ODNI), Intelligence Advanced Research Projects Activity (IARPA), through Army Research Office grant W911NF-09-1-0375.
\end{acknowledgments}


\begin{thebibliography}{10}

\bibitem{galiautdinov2012resonator}
A.~Galiautdinov, A.~Korotkov, and J.~Martinis,
\newblock Phys. Rev. A {\bf 85}, 042321 (2012).

\bibitem{mariantoni2011implementing}
M.~Mariantoni {\em et~al.},
\newblock Science {\bf 334}, 61 (2011).

\bibitem{zmuidzinas2003broadband}
P.~Day, H.~LeDuc, B.~Mazin, A.~Vayonakis, and J.~Zmuidzinas,
\newblock Nature {\bf 425}, 817 (2003).

\bibitem{mazin2012superconducting}
B.~Mazin {\em et~al.},
\newblock Opt. Express {\bf 20}, 1503 (2012).

\bibitem{phillips1999two}
W.~Phillips,
\newblock Rep. Prog. Phys. {\bf 50}, 1657 (1999).

\bibitem{gao2008experimental}
J.~Gao {\em et~al.},
\newblock Appl. Phys. Lett. {\bf 92}, 152505 (2008).

\bibitem{barends2010reduced}
R.~Barends {\em et~al.},
\newblock Appl. Phys. Lett. {\bf 97}, 033507 (2010).

\bibitem{gao2008semiempirical}
J.~Gao {\em et~al.},
\newblock Appl. Phys. Lett. {\bf 92}, 212504 (2008).

\bibitem{gao2011strongly}
J.~Gao {\em et~al.},
\newblock Appl. Phys. Lett. {\bf 98}, 232508 (2011).

\bibitem{martinis2005decoherence}
J.~Martinis {\em et~al.},
\newblock Phys. Rev. Lett. {\bf 95}, 210503 (2005).

\bibitem{kline2008josephson}
J.~Kline, H.~Wang, S.~Oh, J.~Martinis, and D.~Pappas,
\newblock Superconductor Science and Technology {\bf 22}, 015004 (2008).

\bibitem{vissers2012reduced}
M.~Vissers, J.~Kline, J.~Gao, D.~Wisbey, and D.~Pappas,
\newblock Appl. Phys. Lett. {\bf 100}, 082602 (2012).

\bibitem{corcoles2011protecting}
A.~C{\'o}rcoles {\em et~al.},
\newblock Appl. Phys. Lett. {\bf 99}, 181906 (2011).

\bibitem{megrant2012planar}
A.~Megrant {\em et~al.},
\newblock Appl. Phys. Lett. {\bf 100}, 113510 (2012).

\bibitem{barends2011minimizing}
R.~Barends {\em et~al.},
\newblock Appl. Phys. Lett. {\bf 99}, 113507 (2011).

\bibitem{barends2009noise}
R.~Barends {\em et~al.},
\newblock Applied Superconductivity, IEEE Transactions on {\bf 19}, 936 (2009).

\bibitem{Note1}
For simplicity, we have ignore thermal effects from a $\protect \qopname \relax
  o{tanh}(\hbar \omega / 2 k_B T)$ term, as well as negative frequency terms
  from the counter rotating response.

\bibitem{wenner2011surface}
J.~Wenner {\em et~al.},
\newblock Appl. Phys. Lett. {\bf 99}, 113513 (2011).

\bibitem{shalibo2010lifetime}
Y.~Shalibo {\em et~al.},
\newblock Phys. Rev. Lett. {\bf 105}, 177001 (2010).

\bibitem{wang2009improving}
H.~Wang {\em et~al.},
\newblock Appl. Phys. Lett. {\bf 95}, 233508 (2009).

\end{thebibliography}
\end{document}